\documentclass[10pt,a4paper]{article}
\usepackage{xurl}
\usepackage{singlecolumn}
\usepackage{amsthm,mathtools}

\numberwithin{equation}{section}

\theoremstyle{definition}
\newtheorem{theorem}{Theorem}[section]
\newtheorem{lemma}[theorem]{Lemma}
\newtheorem{proposition}[theorem]{Proposition}
\newtheorem{corollary}[theorem]{Corollary}
\theoremstyle{remark}
\newtheorem{remark}[theorem]{Remark}

\newcommand{\R}{\mathbb{R}}
\newcommand{\C}{\mathbb{C}}
\newcommand{\Q}{\mathbb{Q}}
\newcommand{\ii}{\mathrm{i}}
\newcommand{\Ree}{\operatorname{Re}}
\newcommand{\pair}{\mathcal{P}}
\newcommand{\bx}{\mathcal{K}}

\title{\vspace{-1.2em}\bfseries On Robust Strictly Positive Real Synthesis for\\High-Order Interval Polynomials\vspace{-0.4em}}
\author{\normalsize Tiancheng Xu\textsuperscript{1}\quad Guowei Dou\textsuperscript{1}\quad Xiang Ji\textsuperscript{1}\quad ZhengTai Xiong\textsuperscript{1}\quad Long Wang\textsuperscript{2,3}\quad Wensheng Yu\textsuperscript{1}}
\date{\vspace{-1.2em}}
\renewcommand{\runningtitle}{Robust SPR Synthesis for High-Order Interval Polynomials}
\renewcommand{\reporttype}{}
\hypersetup{
  pdftitle={On Robust Strictly Positive Real Synthesis for High-Order Interval Polynomials},
  pdfauthor={Tiancheng Xu, Guowei Dou, Xiang Ji, ZhengTai Xiong, Long Wang, and Wensheng Yu}
}

\begin{document}
\maketitle
\footnotetext[1]{School of Electronic Engineering, Beijing University of Posts and Telecommunications, Beijing 100876, China.}
\footnotetext[2]{Center for Systems and Control, College of Engineering, Peking University, Beijing 100871, China.}
\footnotetext[3]{Center for Multi-Agent Research, Institute for Artificial Intelligence, Peking University, Beijing 100871, China.}

\begin{abstract}
We establish two complementary results on common strictly positive real (SPR) numerators for Hurwitz-stable interval polynomial families. First, for every order $n \ge 10$, we construct a monic Hurwitz-stable interval polynomial family with rational endpoints and positive width in every nonleading coefficient such that every nonzero real numerator of degree at most $n$ produces a strictly negative real part at some member and some positive frequency. In particular, robust Hurwitz stability alone does not guarantee equal-degree SPR synthesis in high orders. Second, for every order $n \ge 1$, we prove that a common real SPR numerator of exact degree $n$ always exists whenever the lower endpoint of each interior coefficient interval is at most $10^{-5}$ times its upper endpoint, while the constant and leading coefficients vary over arbitrary compact positive intervals.
\end{abstract}

\section{Introduction}\label{sec:intro}

Strictly positive real (SPR) transfer functions play a central role in passivity analysis, absolute stability, adaptive control, and output-error system identification~\cite{DasguptaBhagwat1987,YuWang2008}. For a single Hurwitz-stable denominator $a(s)$, choosing $b(s) = a(s)$ trivially yields a transfer function with strictly positive real part on the imaginary axis. When the denominator varies over an uncertain family, however, robust SPR synthesis requires a \emph{single} fixed real numerator $b(s)$ of prescribed degree such that $\Ree[b(\ii\omega)/a(\ii\omega)] > 0$ holds simultaneously for every admissible denominator $a$ and every real frequency $\omega$.

For real interval polynomial families, in which each coefficient varies independently in a closed interval, Kharitonov's theorem~\cite{Kharitonov1978} reduces robust Hurwitz stability to four vertex polynomials. For a \emph{prescribed} numerator, checking the SPR condition across the entire family likewise reduces to these four vertices~\cite{AndersonEtAl1990,WangYu2001}. Synthesis is fundamentally harder because the same numerator must satisfy all four vertex inequalities at all frequencies simultaneously. For low-order Hurwitz-stable interval families ($n \le 4$), Anderson et al.~\cite{AndersonEtAl1990}, Betser and Zeheb~\cite{BetserZeheb1993}, and Yu and Wang~\cite{YuWangFourth2001,WangYu2001} established the existence of a common SPR numerator. For convex combinations of two polynomials (segments), common-numerator synthesis is possible in arbitrary order~\cite{YuWangAckermann2004}. Yet an interval box allows all coefficients to vary independently, so the segment theorem does not settle whether robust Hurwitz stability suffices for interval families of arbitrary order.

In this paper, we resolve the high-order question negatively while identifying a uniform structural regime in which synthesis succeeds at every order:
\begin{enumerate}[label=(\arabic*),leftmargin=2.2em,itemsep=2pt]
  \item \emph{Counterexamples for all $n \ge 10$ (Theorem~\ref{thm:negative}).} Robust Hurwitz stability does not guarantee a common degree-$n$ SPR numerator when $n \ge 10$. At order ten, an explicit rational monic Hurwitz box with positive width in all ten nonleading coefficients admits twelve vertex--frequency evaluation rows of full column rank $11$ with a strictly positive linear dependence and a negative extra moment; this rules out every nonzero numerator of degree at most $10$ (even weakly). Inductive low-end degree lifting $a(s) \mapsto s a(s) + \eta$ propagates this finite obstruction to every order $n \ge 10$.
  \item \emph{All-order synthesis under wide interior variation (Theorem~\ref{thm:positive}).} Conversely, in every order $n \ge 1$, if each interior coefficient interval satisfies $l_j/u_j \le 10^{-5}$ ($1 \le j < n$), then whole-box Hurwitz stability forces successive characteristic frequency scales apart. Decoupling the resulting $n$ frequency bands allows a single degree-$n$ numerator to be constructed via Brouwer's fixed-point theorem (for small stability margins) and bandwise convexity.
\end{enumerate}

The remainder of the paper is organized as follows. Section~\ref{sec:problem} formulates the interval synthesis problem, recalls the four-vertex reduction, and states Theorems~\ref{thm:negative} and~\ref{thm:positive}. Section~\ref{sec:negative} establishes the order-ten finite obstruction and its inductive propagation to all $n \ge 10$. Section~\ref{sec:positive} proves the all-order existence theorem under the interior endpoint-ratio condition. Section~\ref{sec:conclusion} concludes with remarks on open orders and sharp thresholds, and Appendix~\ref{app:certificate} records the exact rational data for the order-ten certificate.

\section{Problem Formulation and Main Results}\label{sec:problem}

Write polynomial coefficients in ascending order and consider the $n$th-order interval polynomial box
\begin{equation}\label{eq:box}
  \bx = \left\{ a(s) = \sum_{j=0}^{n} a_j s^j : l_j \le a_j \le u_j,\ 0 \le j \le n \right\},
  \qquad 0 < l_j \le u_j < \infty.
\end{equation}
A real polynomial $a \in \R[s]$ is \emph{Hurwitz stable} if all of its complex roots lie in the open left half-plane $\{s \in \C : \Ree s < 0\}$, and the box $\bx$ is \emph{Hurwitz} (or robustly Hurwitz stable) if every member $a \in \bx$ is Hurwitz stable. Following the equal-degree convention in~\cite{YuWang2008}, a real polynomial $b \in \R[s]$ of exact degree $n$ is a \emph{common SPR numerator} for $\bx$ if
\begin{equation}\label{eq:target}
  \Ree\frac{b(\ii\omega)}{a(\ii\omega)} > 0
  \qquad (\forall\,a \in \bx,\ \forall\,\omega \in \R).
\end{equation}

Setting $t = \omega^2 \ge 0$, decompose any real polynomial $a(s) = \sum_{j=0}^n a_j s^j$ into its even and odd parts along the imaginary axis:
\begin{equation}\label{eq:parity}
  a(\ii\omega) = E_a(t) + \ii\omega O_a(t), \qquad
  E_a(t) = \sum_{k \ge 0} a_{2k}(-t)^k, \qquad
  O_a(t) = \sum_{k \ge 0} a_{2k+1}(-t)^k.
\end{equation}
For a candidate numerator $b(s) = \sum_{j=0}^n b_j s^j$, define the real frequency pairing polynomial
\begin{equation}\label{eq:row}
  \pair_{a,b}(t) \coloneqq E_a(t)E_b(t) + t\,O_a(t)O_b(t) = R(a, t) \cdot (b_0, \ldots, b_n),
\end{equation}
where the evaluation row vector $R(a, t) = \bigl(R_0(a,t), \ldots, R_n(a,t)\bigr) \in \R^{n+1}$ has entries
\begin{equation}\label{eq:row-entries}
  R_{2k}(a, t) = (-t)^k E_a(t), \qquad
  R_{2k+1}(a, t) = (-t)^k t\,O_a(t).
\end{equation}
Whenever $a$ is Hurwitz stable, $|a(\ii\omega)|^2 = E_a(\omega^2)^2 + \omega^2 O_a(\omega^2)^2 > 0$, so
\begin{equation}\label{eq:re-ratio}
  \Ree\frac{b(\ii\omega)}{a(\ii\omega)} = \frac{\pair_{a,b}(\omega^2)}{|a(\ii\omega)|^2},
\end{equation}
and the sign of $\Ree[b(\ii\omega)/a(\ii\omega)]$ coincides with the sign of $\pair_{a,b}(\omega^2)$.

\begin{lemma}[Four-vertex reduction~\cite{AndersonEtAl1990,WangYu2001,YuWang2008}]\label{lem:four-vertex}
For each fixed $t \ge 0$, as $a$ ranges over $\bx$, the pair $\bigl(E_a(t), O_a(t)\bigr)$ sweeps a closed axis-aligned rectangle whose four corners are realized by four Kharitonov vertex polynomials in $\bx$ independent of $t$. Consequently, for any fixed $b \in \R[s]$ and $t \ge 0$, $\pair_{a,b}(t) > 0$ (respectively, $\ge 0$) holds for all $a \in \bx$ if and only if it holds at these four vertices.
\end{lemma}

We now state the two main theorems of the paper.

\begin{theorem}[Counterexamples for all $n \ge 10$]\label{thm:negative}
For every integer $n \ge 10$, there exists a Hurwitz box $\bx$ of the form~\eqref{eq:box} with $l_n = u_n = 1$ and rational endpoints $l_j < u_j$ for every $0 \le j < n$ such that, for every nonzero real polynomial $b \in \R[s]$ of degree at most $n$, there exist $a \in \bx$ and $\omega > 0$ satisfying
\[
  \Ree\frac{b(\ii\omega)}{a(\ii\omega)} < 0.
\]
At order $n = 10$, twelve fixed tests $(a_m, \omega_m)$ on four Kharitonov vertices suffice for all such $b$.
\end{theorem}

\begin{theorem}[All-order synthesis under wide interior intervals]\label{thm:positive}
Let $n \ge 1$, and let $\bx$ be a Hurwitz box of the form~\eqref{eq:box}. If the interior coefficient intervals satisfy
\begin{equation}\label{eq:width}
  \frac{l_j}{u_j} \le 10^{-5} \qquad (1 \le j < n),
\end{equation}
then there exists a real polynomial $b \in \R[s]$ of exact degree $n$ satisfying~\eqref{eq:target}.
\end{theorem}

\begin{remark}[Two contrasting regimes]\label{rem:regimes}
The threshold~\eqref{eq:width} in Theorem~\ref{thm:positive} is uniform in both the order $n$ and the stability margin of $\bx$, and imposes no width restriction on $a_0$ or $a_n$. Note the direction of the inequality:~\eqref{eq:width} requires each interior coefficient to vary across a factor of at least $10^5$. Under the joint hypothesis that the \emph{entire} box is Hurwitz stable, such wide variation forces adjacent characteristic frequency scales apart, decoupling the synthesis constraints across frequency bands. By contrast, the order-ten counterexample in Theorem~\ref{thm:negative} has several interior ratios $l_j/u_j = 1 - 10^{-300}$ close to $1$; the obstruction arises because all frequency constraints must be met by a single numerator, and the twelve evaluation rows on four vertices have full column rank $11$ together with a strictly positive linear dependence, forcing every nonzero numerator to take a strictly negative value on at least one row.
\end{remark}

\section{Counterexamples for Orders \texorpdfstring{$n \ge 10$}{n >= 10}}\label{sec:negative}

In this section we prove Theorem~\ref{thm:negative}. We begin with a general finite-matrix criterion that certifies the nonexistence of any nonzero weak numerator (Lemma~\ref{lem:obstruction-criterion}), apply it to the explicit order-ten box (Proposition~\ref{prop:seed} and Corollary~\ref{cor:order-ten}), and then lift the obstruction inductively from order $d$ to $d+1$ (Lemma~\ref{lem:lift}).

\subsection{Finite Obstruction Criterion and the Order-Ten Counterexample}

\begin{lemma}[Finite obstruction criterion]\label{lem:obstruction-criterion}
Let $\bx$ be a Hurwitz box of degree $d \ge 1$. Suppose there exist finitely many test pairs $(a_m, t_m) \in \bx \times (0, \infty)$ ($1 \le m \le M$) and strictly positive weights $\lambda_1, \ldots, \lambda_M > 0$ such that the $M \times (d+1)$ matrix $R$ with rows $R(a_m, t_m)$ has full column rank $d+1$ and satisfies
\begin{equation}\label{eq:zero-dep}
  \sum_{m=1}^{M} \lambda_m R(a_m, t_m) = 0.
\end{equation}
Then for every nonzero real polynomial $b \in \R[s]$ of degree at most $d$, there exists an index $m \in \{1, \ldots, M\}$ such that
\[
  \Ree\frac{b(\ii\sqrt{t_m})}{a_m(\ii\sqrt{t_m})} < 0.
\]
\end{lemma}

\begin{proof}
Identify $b(s) = \sum_{j=0}^d b_j s^j$ with its nonzero coefficient vector $\mathbf{b} = (b_0, \ldots, b_d)^{\mathsf{T}} \in \R^{d+1}$. Since $R$ has full column rank $d+1$, the vector $R\mathbf{b} \in \R^M$ is nonzero. On the other hand,~\eqref{eq:zero-dep} gives $\sum_{m=1}^M \lambda_m (R\mathbf{b})_m = 0$ with all $\lambda_m > 0$. Hence at least one entry $(R\mathbf{b})_m = \pair_{a_m, b}(t_m)$ must be strictly negative. Because $a_m \in \bx$ is Hurwitz stable, $|a_m(\ii\sqrt{t_m})|^2 > 0$, and~\eqref{eq:re-ratio} yields $\Ree[b(\ii\sqrt{t_m})/a_m(\ii\sqrt{t_m})] < 0$.
\end{proof}

\begin{proposition}[Order-ten seed obstruction]\label{prop:seed}
There exist a rational monic Hurwitz box $\bx_{10}$ of degree ten with $l_j < u_j$ for all $0 \le j \le 9$, twelve test pairs $(a_m, t_m)$ ($1 \le m \le 12$) on four vertices of $\bx_{10}$ with $t_m > 0$, and strictly positive weights $\lambda_1, \ldots, \lambda_{12} > 0$ such that the $12 \times 11$ matrix $R$ with rows $R(a_m, t_m)$ has full column rank $11$ and satisfies
\begin{equation}\label{eq:moments}
  \sum_{m=1}^{12} \lambda_m R(a_m, t_m) = 0, \qquad
  \sum_{m=1}^{12} \lambda_m \bigl[-O_{a_m}(t_m)\bigr] < 0.
\end{equation}
\end{proposition}

\begin{proof}
The explicit rational endpoints of $\bx_{10}$, the root-bracketing intervals, the twelve vertex--frequency pairs, and the rational determinant/linear-system checks are given in Appendix~\ref{app:certificate}.

For Hurwitz stability of every $a \in \bx_{10}$, four disjoint intervals $[L_h, H_h]$ ($h = 1, 2, 3, 4$) each bracket a strict sign change of $O_a(t)$ uniformly over $\bx_{10}$, yielding four distinct positive roots of the quartic $O_a(t)$. At the boundary points $L_h, H_h$ together with $t = 0$ and $t \to \infty$, the signs of $E_a(t)$ force one root of the quintic $E_a(t)$ in each of the five complementary open gaps $(0, L_1), (H_1, L_2), (H_2, L_3), (H_3, L_4), (H_4, \infty)$. Since $\deg E_a = 5$ and $\deg O_a = 4$, these roots are simple and strictly interlaced, and $E_a(0), O_a(0) > 0$. By the Hermite--Biehler theorem~\cite{Holtz2003}, every $a \in \bx_{10}$ is Hurwitz stable.

For the matrix conditions~\eqref{eq:moments}, exact rational elimination shows that the first eleven rows $R(a_1, t_1),$ $\ldots,$ $R(a_{11}, t_{11})$ are linearly independent, so $R$ has rank $11$. Expressing $-R(a_{12}, t_{12})$ in the basis of the first eleven rows yields eleven strictly positive coordinates $\lambda_1, \ldots, \lambda_{11} > 0$. Setting $\lambda_{12} = 1$ gives $\sum_{m=1}^{12} \lambda_m R(a_m, t_m) = 0$, and direct evaluation of the same weights verifies $\sum_{m=1}^{12} \lambda_m[-O_{a_m}(t_m)] < 0$.
\end{proof}

Combining Lemma~\ref{lem:obstruction-criterion} with the first identity in Proposition~\ref{prop:seed} immediately establishes the order-ten case of Theorem~\ref{thm:negative}; the second inequality in~\eqref{eq:moments} supplies the extra sign needed for degree lifting.

\begin{corollary}[Order-ten nonexistence]\label{cor:order-ten}
For the rational monic Hurwitz box $\bx_{10}$ and the twelve fixed vertex--frequency pairs $(a_m, t_m)$ in Proposition~\ref{prop:seed}, every nonzero real polynomial $b \in \R[s]$ of degree at most $10$ satisfies $\Ree[b(\ii\sqrt{t_m})/a_m(\ii\sqrt{t_m})] < 0$ for at least one $m \in \{1, \ldots, 12\}$.
\end{corollary}

\subsection{Inductive Degree Lifting and Proof of Theorem~\ref{thm:negative}}

\begin{lemma}[Low-end degree lifting]\label{lem:lift}
Suppose a compact monic Hurwitz box $\bx_d$ of degree $d \ge 1$ admits finitely many positive-frequency tests $(a_m, t_m)$ ($1 \le m \le M$) whose evaluation matrix has full column rank $d+1$ and whose positive weights $\lambda_m > 0$ satisfy~\eqref{eq:moments} (with sum over $1 \le m \le M$). Then there exists a rational monic Hurwitz box $\bx_{d+1}$ of degree $d+1$, with every nonleading coefficient interval of positive width, admitting $M+1$ positive-frequency tests with full column rank $d+2$ and positive weights satisfying~\eqref{eq:moments}.
\end{lemma}

\begin{proof}
We construct $\bx_{d+1}$ in three steps.

\emph{Step 1: Hurwitz stability of the shifted family.}
For a parameter $\eta > 0$, replace each $a \in \bx_d$ by the monic degree-$(d+1)$ polynomial
\[
  A(s) = s\,a(s) + \eta.
\]
These polynomials form a compact monic box (initially with constant coefficient fixed at $\eta$). At $\eta = 0$, every $s\,a(s)$ has a simple root at $s = 0$ because $a_0 \ge l_0 > 0$, while its remaining $d$ roots lie uniformly inside the open left half-plane by compactness of $\bx_d$. Differentiating $A(s(\eta)) = 0$ at $\eta = 0$ gives $s'(0) = -1/a_0 < 0$. By the implicit function theorem on the compact box $\bx_d$, the continuation of the zero root admits a uniform expansion $-\eta/a_0 + O(\eta^2)$, which is strictly negative for all sufficiently small $\eta > 0$, while the other $d$ roots remain in the open left half-plane. Thus the lifted box is Hurwitz for all sufficiently small $\eta > 0$.

\emph{Step 2: Continuation of the positive dependence and an added test near $t = 0$.}
From~\eqref{eq:parity}, the even and odd parts of $A(s) = s\,a(s) + \eta$ are
\begin{equation}\label{eq:lifted-EO}
  E_A(t) = \eta - t\,O_a(t), \qquad O_A(t) = E_a(t).
\end{equation}
Comparing~\eqref{eq:row-entries} at degree $d+1$ and degree $d$, when $\eta = 0$ the lifted row $R^{(d+1)}(A_m, t_m) \in \R^{d+2}$ satisfies
\[
  R^{(d+1)}_0(A_m, t_m) = -t_m O_{a_m}(t_m), \qquad
  R^{(d+1)}_{j+1}(A_m, t_m) = t_m R^{(d)}_j(a_m, t_m) \quad (0 \le j \le d).
\]
Consider the rescaled weights $\tilde\lambda_m(0) = \lambda_m / t_m > 0$. At $\eta = 0$, they annihilate columns $1, \ldots, d+1$ of the lifted matrix and yield a strictly negative sum in column $0$:
\[
  S(0) \coloneqq \sum_{m=1}^{M} \tilde\lambda_m(0)\bigl[-t_m O_{a_m}(t_m)\bigr]
  = \sum_{m=1}^{M} \lambda_m\bigl[-O_{a_m}(t_m)\bigr] < 0.
\]
Since columns $1, \ldots, d+1$ at $\eta = 0$ have rank $d+1$, choose $d+1$ linearly independent rows of this submatrix. Keeping the remaining $M - (d+1)$ weights fixed and solving the $(d+1) \times (d+1)$ linear system on this nonsingular minor continues $\tilde\lambda_m(0)$ continuously to strictly positive weights $\tilde\lambda_m(\eta) > 0$ for small $\eta > 0$ that still annihilate columns $1, \ldots, d+1$, while their column-$0$ sum satisfies $S(\eta) \to S(0) < 0$.

Now pick any member $a_* \in \bx_d$ and add an $(M+1)$-th test on $A_*(s) = s\,a_*(s) + \eta$, temporarily at $t = 0$. By~\eqref{eq:lifted-EO}, its evaluation row is $(\eta, 0, \ldots, 0)$ and its extra moment is $-O_{A_*}(0) = -E_{a_*}(0) = -a_*(0) < 0$. Assign it the positive weight $\mu(\eta) = -S(\eta)/\eta > 0$. Then:
\begin{itemize}[leftmargin=1.8em,itemsep=1pt]
  \item all $d+2$ columns of the $(M+1) \times (d+2)$ matrix have weighted sum zero;
  \item the matrix has full column rank $d+2$ (the new row is nonzero only in column $0$, while the first $M$ rows have rank $d+1$ on columns $1, \ldots, d+1$);
  \item as $\eta \downarrow 0$, the extra moment from the first $M$ tests remains bounded, whereas the new test contributes $-\mu(\eta)a_*(0) = \frac{S(\eta)}{\eta}a_*(0) \to -\infty$.
\end{itemize}
Fix a sufficiently small $\eta > 0$ so that the lifted box is Hurwitz and the total extra moment is strictly negative. Perturbing the $(M+1)$-th node from $t = 0$ to a sufficiently small $t_{M+1} > 0$ and solving continuously on the enlarged $(d+2) \times (d+2)$ nonsingular minor preserves full column rank $d+2$, positivity of all $M+1$ weights, and strict negativity of the extra moment.

\emph{Step 3: Rational full-width enlargement.}
By openness of the Hurwitz property and compactness of the lifted box, we may slightly enlarge each nonleading coefficient interval (including the constant coefficient $[\eta, \eta]$) to a rational closed interval of positive width while keeping the entire box Hurwitz stable. Every tested polynomial $A_m$ remains inside the enlarged rational box $\bx_{d+1}$, completing the induction step.
\end{proof}

\begin{proof}[Proof of Theorem~\ref{thm:negative}]
For $n = 10$, the assertion is Corollary~\ref{cor:order-ten}. For any $n > 10$, applying Lemma~\ref{lem:lift} inductively $n - 10$ times starting from Proposition~\ref{prop:seed} produces a rational monic Hurwitz box $\bx_n$ with positive width in all nonleading coefficients and a finite positive-frequency test set satisfying the hypotheses of Lemma~\ref{lem:obstruction-criterion}, which yields the stated inequality for every nonzero $b \in \R[s]$ of degree at most $n$.
\end{proof}

\section{All-Order Synthesis Under the Endpoint-Ratio Condition}\label{sec:positive}

We now prove Theorem~\ref{thm:positive}. Since the cases $n = 1$ and $n = 2$ are elementary (handled at the end of Section~\ref{subsec:global-pos}), we assume throughout Sections~\ref{subsec:scale-sep}--\ref{subsec:bands} that $n \ge 3$ and that $\bx$ is a Hurwitz box satisfying~\eqref{eq:width}. Define the coefficient ratios, maximum interior ratio, and four-coefficient stability ratios by
\begin{equation}\label{eq:xi-D-def}
  \xi_j \coloneqq \frac{l_j}{u_j} \quad (0 \le j \le n), \qquad
  \epsilon \coloneqq \max_{1 \le j < n} \xi_j \le 10^{-5}, \qquad
  D_j \coloneqq \frac{l_j l_{j+1}}{u_{j-1} u_{j+2}} \quad (1 \le j \le n-2).
\end{equation}
A key structural feature of the proof is the distinction between the \emph{small-margin regime} $1 < D_j \le 4$ (where the stability margin can be arbitrarily small, so crude approximations can lose positivity) and the \emph{large-margin regime} $D_j > 4$ (where direct coefficient dominance suffices).

\subsection{Four-Coefficient Ratios and Baseline Scale Separation}\label{subsec:scale-sep}

\begin{lemma}[Four-coefficient stability inequality]\label{lem:Dj-gt-one}
Every Hurwitz-stable polynomial $a(s) = \sum_{j=0}^n a_j s^j$ satisfies $a_j a_{j+1} > a_{j-1} a_{j+2}$ for all $1 \le j \le n-2$. Consequently, for the Hurwitz box $\bx$, every four-coefficient ratio satisfies $D_j > 1$ ($1 \le j \le n-2$).
\end{lemma}

\begin{proof}
Set $P(x) = E_a(-x)$ and $Q(x) = O_a(-x)$. By Hermite--Biehler strict interlacing~\cite{Holtz2003}, the roots of $P$ and $x Q$ are simple, nonpositive, and strictly alternating, so both $Q/P$ and $P/(xQ)$ admit partial-fraction expansions of the form $c + \sum w/(x+\alpha)$ with $w > 0$ and $\alpha \ge 0$ (in $P/(xQ)$, the residue at $x = 0$ is $P(0)/Q(0) = a_0/a_1 > 0$). For either quotient $A/B$ and each pole $-\alpha$, set $H = B/(x+\alpha)$ (taking out-of-range coefficients to be zero). Comparing coefficients in $A = cB + \sum w H$ yields
\[
  A_k B_{k+1} - B_k A_{k+1} = \sum w \bigl(H_k^2 - H_{k-1}H_{k+1}\bigr).
\]
Since all roots of $H$ are real and nonpositive, Newton's inequalities give $H_k^2 - H_{k-1}H_{k+1} \ge 0$, with strict inequality whenever $H_k > 0$. For $Q/P$, every $H$ has positive coefficients throughout its degree range; for $P/(xQ)$, the pole at zero alone contributes $H = Q$. Reading these two sets of inequalities in alternating index order gives $a_j a_{j+1} > a_{j-1} a_{j+2}$ for every $1 \le j \le n-2$. Choosing $a_j = l_j$, $a_{j+1} = l_{j+1}$, $a_{j-1} = u_{j-1}$, and $a_{j+2} = u_{j+2}$ in $\bx$ yields $D_j > 1$.
\end{proof}

\begin{lemma}[Baseline scales and geometric separation]\label{lem:separation}
There exist positive baseline coefficients $\beta_0, \ldots, \beta_n > 0$ satisfying
\begin{equation}\label{eq:betaineq}
  \frac{\beta_j}{\beta_{j+1}} \le \frac{l_{j+1}}{u_{j+2}} \quad (0 \le j \le n-2), \qquad
  \frac{\beta_{j+1}}{\beta_j} \le \frac{l_j}{u_{j-1}} \quad (1 \le j \le n-1).
\end{equation}
Writing $W_j \coloneqq u_j / u_{j+2}$ ($0 \le j \le n-2$) and $\tau_j \coloneqq \beta_{j-1} / \beta_{j+1}$ ($1 \le j \le n-1$), we have the geometric scale separation
\begin{equation}\label{eq:scales}
  \frac{W_j}{W_{j+1}} \le \epsilon^2 \quad (0 \le j \le n-3), \qquad
  \frac{\tau_j}{\tau_{j+1}} \le \epsilon^2 \quad (1 \le j \le n-2).
\end{equation}
\end{lemma}

\begin{proof}
Fix $\delta = 1/16$, set $\beta_1 = 1$ and $\beta_0 = \delta l_1 / u_2$, and define the interior ratios inductively by
\begin{equation}\label{eq:betas}
  \frac{\beta_{i+1}}{\beta_i} =
  \begin{cases}
    \dfrac{u_{i+2}}{l_{i+1}}, & 1 < D_i \le 4, \\[8pt]
    \sqrt{\dfrac{u_{i+2} l_i}{l_{i+1} u_{i-1}}}, & D_i > 4,
  \end{cases}
  \qquad (1 \le i \le n-2),
\end{equation}
and finally $\beta_n = \delta l_{n-1}\beta_{n-1}/u_{n-2}$. For each $1 \le i \le n-2$, introduce the dimensionless quantities
\begin{equation}\label{eq:AiBi}
  A_i \coloneqq \frac{u_{i+2}\beta_i}{l_{i+1}\beta_{i+1}}, \qquad
  B_i \coloneqq \frac{u_{i-1}\beta_{i+1}}{l_i\beta_i}.
\end{equation}
By~\eqref{eq:betas} and Lemma~\ref{lem:Dj-gt-one}, $(A_i, B_i) = (1, 1/D_i)$ when $1 < D_i \le 4$, and $A_i = B_i = 1/\sqrt{D_i} < 1/2$ when $D_i > 4$. Together with $\beta_0/\beta_1 = \delta l_1/u_2 < l_1/u_2$ and $\beta_n/\beta_{n-1} = \delta l_{n-1}/u_{n-2} < l_{n-1}/u_{n-2}$, this proves~\eqref{eq:betaineq}. Finally,
\[
  \frac{W_j}{W_{j+1}} = \frac{u_j u_{j+3}}{u_{j+1} u_{j+2}} = \frac{\xi_{j+1}\xi_{j+2}}{D_{j+1}} \le \epsilon^2,
\]
and multiplying $\beta_{j-1}/\beta_j \le l_j/u_{j+1}$ with $\beta_{j+2}/\beta_{j+1} \le l_{j+1}/u_j$ (from the second inequality in~\eqref{eq:betaineq} at index $j+1$) gives $\tau_j/\tau_{j+1} = (\beta_{j-1}/\beta_j)(\beta_{j+2}/\beta_{j+1}) \le (l_j/u_{j+1})(l_{j+1}/u_j) = \xi_j \xi_{j+1} \le \epsilon^2$, establishing~\eqref{eq:scales}.
\end{proof}

\subsection{Stationary-Point Alignment in the Small-Margin Regime}\label{subsec:stationary}

For each $0 \le i < n$, let $a^{(i)} \in \bx$ denote the Kharitonov vertex taking lower endpoints $l_j$ at indices $j \equiv i, i+1 \pmod{4}$ and upper endpoints $u_j$ at all other indices. As shown in Lemma~\ref{lem:bands} below, $a^{(i)}$ will be the minimizing vertex of $\pair_{a,b}(t)$ on the $i$th frequency band.

To handle indices with $1 < D_i \le 4$, we perturb the baseline coefficients $\beta_j$ by positive multipliers $z_j$ ($b_j = \beta_j z_j$) with ratios $x_i = z_{i+1}/z_i$ chosen via Brouwer's fixed-point theorem so that the normalized pairing on band $i$ has a stationary point precisely at a zero of the rotated real part of $a^{(i)}(\ii\omega)$.

\begin{proposition}\label{prop:stationary}
There exist positive coefficients $b_j = \beta_j z_j$ ($0 \le j \le n$) with $z_0 = z_1 = 1$ and $z_n = z_{n-1}$ such that the interior ratios $x_i \coloneqq z_{i+1}/z_i$ ($1 \le i \le n-2$) satisfy $|x_i - 1| \le 4000\epsilon$, with $x_i = 1$ whenever $D_i > 4$. Moreover, for each $1 \le i \le n-2$ with $D_i \le 4$, setting $T_i \coloneqq l_i / u_{i+2}$, the normalized pairing
\begin{equation}\label{eq:normalized}
  H_i(v) \coloneqq \frac{\pair_{a^{(i)}, b}(T_i v)}{b_i l_i (T_i v)^i}
\end{equation}
has a stationary point $v_i \in [3/4, 5/4]$ with $H_i(v_i) > 0$.
\end{proposition}

\begin{proof}
Let $x = (x_1, \ldots, x_{n-2})$ range over the cube $[7/8, 9/8]^{n-2}$ and recover $z_0, \ldots, z_n$ from $z_0 = z_1 = 1$, $z_{i+1} = x_i z_i$, and $z_n = z_{n-1}$. Adjacent $z_j$ differ by at most a factor of $8/7$ in either direction.

Fix an index $i$ with $1 < D_i \le 4$, and set $t = T_i v = \omega^2$ ($\omega > 0$) for $v \in [1/2, 2]$. Divide both $a^{(i)}(\ii\omega)$ and $b(\ii\omega)$ by $(\ii\omega)^i$, and express their real and imaginary parts in units of $(l_i, \omega l_{i+1})$ for $a^{(i)}$ and $(b_i, \omega \beta_{i+1} z_i)$ for $b$. Using $a^{(i)}_i = l_i$, $a^{(i)}_{i+1} = l_{i+1}$, $a^{(i)}_{i+2} = u_{i+2}$, $a^{(i)}_{i-1} = u_{i-1}$, and $T_i = l_i / u_{i+2}$, these four normalized components take the form
\begin{equation}\label{eq:local}
  R_a(v) = 1 - v + e_1(v), \quad
  I_a(v) = 1 - (D_i v)^{-1} + e_2(v), \qquad
  R_b(v) = 1 + e_3(v), \quad
  I_b(v) = x_i + e_4(v).
\end{equation}
Define the $C^1$ norm $\|f\|_1 \coloneqq \sup_{v \in [1/2, 2]} |f(v)| + \sup_{v \in [1/2, 2]} |f'(v)|$. We claim that $\|e_1\|_1, \|e_2\|_1 < 64\epsilon^2$ and $\|e_3\|_1, \|e_4\|_1 < 64\epsilon$. Indeed:
\begin{itemize}[leftmargin=1.8em,itemsep=2pt]
  \item In $R_a$, the nearest omitted lower term at $v = 1$ has size at most $u_{i-2}u_{i+2}/l_i^2 = \xi_{i-1}\xi_{i+1}/(D_{i-1}D_i) \le \epsilon^2$, and the nearest upper term has size at most $T_i / W_{i+2} \le \epsilon^2$ (since $T_i = \xi_i W_i$ and $W_i / W_{i+2} \le \epsilon^4$).
  \item In $I_a$, the nearest omitted lower term at $v = 1$ (present when $i \ge 3$) has magnitude $(\xi_{i-3}/D_i)\,W_{i-3}/T_i$, which is bounded above by $W_{i-3}/T_i \le \epsilon^2$ (using $W_{i-1}/T_i = \xi_{i+1}/D_i$ and~\eqref{eq:scales}), and the nearest omitted upper term has magnitude at most $\xi_i \xi_{i+2}/D_{i+1} \le \epsilon^2$.
  \item By~\eqref{eq:scales}, each subsequent step outwards in $R_a$ or $I_a$ introduces an additional factor of at most $\epsilon^2$ (out-of-range terms are absent).
  \item In $R_b$ and $I_b$,~\eqref{eq:betaineq} and the first branch of~\eqref{eq:betas} give $\tau_i / T_i \le \epsilon$ and $T_i / \tau_{i+1} \le \epsilon$. Accounting for the factor $(8/7)^2$ from adjacent $z$-ratios, the nearest omitted terms at $v = 1$ cost at most $3\epsilon$, and each subsequent parity step costs at most $2\epsilon^2$.
\end{itemize}
Bounding $v^k$ and its derivative on $[1/2, 2]$ yields the geometric majorants
\[
  \|e_1\|_1, \|e_2\|_1 \le 8\epsilon^2 \sum_{h \ge 0} (5 + 2h)(2\epsilon^2)^h < 64\epsilon^2, \qquad
  \|e_3\|_1, \|e_4\|_1 \le 12\epsilon \sum_{h \ge 0} (3 + 2h)(4\epsilon^2)^h < 64\epsilon.
\]

Since $R_a(v) = 1 - v + e_1(v)$ with $\|e_1\|_1 < 64\epsilon^2$, $R_a$ has a unique simple zero $v_i \in [3/4, 5/4]$ with $R_a'(v_i) < 0$; note that $v_i$ depends only on the vertex $a^{(i)}$ and not on $x$. At this zero, $I_a(v_i) > 0$: along the positive imaginary axis, the argument of any Hurwitz polynomial is strictly increasing (each root $-p + \ii q$ with $p > 0$ contributes $p/[p^2 + (\omega - q)^2] > 0$ to $\frac{\mathrm{d}}{\mathrm{d}\omega}\arg a(\ii\omega)$), and dividing by $(\ii\omega)^i$ merely subtracts a constant; hence at any crossing where the real part decreases through zero, the imaginary part must be strictly positive.

Using $\beta_{i+1}/\beta_i = u_{i+2}/l_{i+1} = l_i/(l_{i+1} T_i)$ from~\eqref{eq:betas}, the normalized pairing~\eqref{eq:normalized} equals
\[
  H_i(v) = R_a(v) R_b(v) + v\,I_a(v) I_b(v).
\]
Its principal part $(1 - v) + (v - D_i^{-1})x_i$ has derivative $x_i - 1$. Using the submultiplicativity of $\|\cdot\|_1$ (with bounds $2, 5, 1, 2$ for the four principal components and $3$ for $v$ on $[1/2, 2]$), we obtain
\[
  \bigl|H_i'(v_i) - (x_i - 1)\bigr| \le 4000\epsilon < \frac{1}{8}.
\]
Define the continuous map $F : [7/8, 9/8]^{n-2} \to \R^{n-2}$ by setting $F_i(x) = H_i'(v_i)$ when $D_i \le 4$, and $F_i(x) = x_i - 1$ when $D_i > 4$. Because $|(x_i - F_i(x)) - 1| \le 4000\epsilon < 1/8$, the map $x \mapsto x - F(x)$ maps the compact convex cube $[7/8, 9/8]^{n-2}$ continuously into itself. By Brouwer's fixed-point theorem, there exists a fixed point $x^*$ at which $F(x^*) = 0$. This gives $x_i^* = 1$ for $D_i > 4$, $|x_i^* - 1| \le 4000\epsilon$ for all $i$, and $H_i'(v_i) = 0$ for all $D_i \le 4$. Finally, at $v_i$, $R_a(v_i) = 0$ and $I_b(v_i) \ge 7/8 - 64\epsilon > 0$, so $H_i(v_i) = v_i I_a(v_i) I_b(v_i) > 0$, using the true sign of $I_a(v_i)$ regardless of how small $D_i - 1$ may be.
\end{proof}

\subsection{Frequency Band Decomposition and Envelope Decay}\label{subsec:bands}

Fix the numerator $b(s) = \sum_{j=0}^n b_j s^j$ constructed in Proposition~\ref{prop:stationary}. Its alternating parity zeros partition $[0, \infty)$ into $n$ frequency bands on each of which a single vertex $a^{(i)}$ minimizes $\pair_{a,b}(t)$.

\begin{lemma}[Frequency band decomposition]\label{lem:bands}
There exist strictly increasing nodes $0 = t_0 < t_1 < \cdots < t_{n-1} < t_n = \infty$ such that, for each $0 \le i < n$, the vertex $a^{(i)}$ minimizes $\pair_{a,b}(t)$ over $\bx$ at every finite $t \in [t_i, t_{i+1}]$. Putting $\sigma_j \coloneqq b_{j-1}/b_{j+1}$ ($1 \le j < n$), we have
\[
  \frac{\sigma_j}{2} < t_j < 2\sigma_j \qquad (1 \le j < n),
\]
and whenever $1 \le i \le n-2$ satisfies $D_i \le 4$, the stationary point satisfies $T_i v_i \in (t_i, t_{i+1})$.
\end{lemma}

\begin{proof}
Since adjacent $z$-ratios differ by at most $(8/7)^2$,~\eqref{eq:scales} implies $\sigma_j / \sigma_{j+1} \le 2\epsilon^2$. Setting $t = \sigma_j v$ ($v \in [1/2, 2]$), the two terms of degrees $j-1$ and $j+1$ in the corresponding parity component ($E_b$ if $j$ is odd, $O_b$ if $j$ is even), normalized by the degree-$(j-1)$ term, give $1 - v$. At $v = 1$, the nearest omitted term on either side has magnitude at most $(2\epsilon^2)^2$, and each subsequent parity step introduces a factor of at most $2\epsilon^2$. The same geometric series bound as in Proposition~\ref{prop:stationary} keeps the $C^1$ error below $128\epsilon^2$, yielding a unique simple zero $t_j \in (\sigma_j/2, 2\sigma_j)$. Meanwhile, the opposite parity component is dominated by its degree-$j$ term (neighboring terms cost at most $2\epsilon^2$ at $v = 1$). Because these $n-1$ intervals are disjoint, they exhaust all zeros of $E_b$ and $O_b$ on $(0, \infty)$.

At $t = 0$, both $E_b(0) = b_0 > 0$ and $O_b(0) = b_1 > 0$. As $t$ crosses $t_1, t_2, \ldots, t_{n-1}$ in turn, the zeros alternate between $E_b$ and $O_b$, starting with $E_b$ at $t_1$. Hence on each band $[t_i, t_{i+1}]$ (for $\omega > 0$), both the real and imaginary parts of $b(\ii\omega)/(\ii\omega)^i$ are nonnegative. Minimizing the pairing $\pair_{a,b}(t)$ in these rotated axes selects the lower endpoints at indices $j \equiv i, i+1 \pmod{4}$ and the upper endpoints elsewhere, which is precisely $a^{(i)}$. Continuity includes the band endpoints and $t = 0$. Finally, for $D_i \le 4$, the bounds $\tau_i/T_i, T_i/\tau_{i+1} \le \epsilon$ from Proposition~\ref{prop:stationary} give $\sigma_i/T_i, T_i/\sigma_{i+1} \le 2\epsilon$, placing $T_i v_i \in [\frac{3}{4}T_i, \frac{5}{4}T_i]$ strictly inside $(t_i, t_{i+1})$.
\end{proof}

For any fixed vertex $a \in \bx$, expand its pairing with $b$ as $\pair_{a,b}(t) = \sum_{k=0}^n c_k t^k$.

\begin{lemma}[Coefficient bounds and tail decay]\label{lem:tails}
For every vertex $a$ of $\bx$ and every $0 \le k \le n$:
\begin{enumerate}[label=(\roman*),leftmargin=2.2em,itemsep=2pt]
  \item $|c_k| \le 4 b_k u_k$, and if $a_k = u_k$, then $c_k \ge (1 - 4\epsilon) b_k u_k$;
  \item on band $i$ ($t \in [t_i, t_{i+1}]$), successive upper envelopes $b_j u_j t^j$ decrease by a factor of at most $q_0 \coloneqq 8\epsilon^3$ when moving upwards from $j \ge i+2$, or downwards from $j \le i-1$.
\end{enumerate}
\end{lemma}

\begin{proof}
Expanding~\eqref{eq:row} yields
\[
  c_k = \sum_{h} (-1)^h b_{k-h} a_{k+h},
\]
summed over all integers $h$ for which both $0 \le k-h \le n$ and $0 \le k+h \le n$. By~\eqref{eq:betaineq} and $z_{k \pm 1}/z_k \le 8/7$, each distance-one term ($h = \pm 1$) has magnitude at most $(8/7) b_k l_k$. Putting $r_j \coloneqq u_{j+1}/u_j$, we have $r_{p+1}/r_{p-1} = \xi_p \xi_{p+1} / D_p \le \epsilon^2$, so $r_{k+h}/r_{k-h} \le \epsilon^{2h}$ for $h \ge 1$. Hence for $h \ge 1$, the ratio of successive envelopes $b_{k-h} u_{k+h}$ satisfies
\[
  \frac{b_{k-h-1} u_{k+h+1}}{b_{k-h} u_{k+h}}
  \le \frac{8}{7}\,\xi_{k-h}\,\frac{r_{k+h}}{r_{k-h}}
  \le 2\epsilon^{2h}.
\]
Reversing coefficient order gives the same bound for $h \le -1$. Summing over $|h| \ge 2$, all terms at distance at least two together cost at most $8\epsilon^2 b_k l_k$. Combining the central term $b_k a_k$ with the distance-one and tail bounds yields (i) (with $c_0 = b_0 a_0$ and $c_n = b_n a_n$ exact).

For (ii), if a high tail exists on band $i$ ($i \le n-2$), Lemma~\ref{lem:bands} gives $t \le t_{i+1} < 2 b_i / b_{i+2}$. For any $j \ge i+2$, applying~\eqref{eq:betaineq} and $W_{j-1} \ge W_{i+1}$ (from~\eqref{eq:scales}) yields
\[
  t\,\frac{b_{j+1} u_{j+1}}{b_j u_j}
  \le 2\left(\frac{8}{7}\right)^3 \frac{l_{i+1} l_{i+2}}{u_{i+2} u_{i+3}}\,\xi_j\,\frac{u_{j+1}}{u_{j-1}}
  \le 8\epsilon^3 = q_0.
\]
The low-tail estimate ($j \le i-1$) follows symmetrically by reversing all coefficient indices and replacing $t$ with $1/t$.
\end{proof}

\subsection{Bandwise Positivity and Proof of Theorem~\ref{thm:positive}}\label{subsec:global-pos}

We verify strict positivity of $\pair_{a^{(i)}, b}(t)$ separately on the interior bands and on the two boundary bands, and then assemble the proof of Theorem~\ref{thm:positive}.

\begin{lemma}[Positivity on interior bands]\label{lem:interior-bands}
For every interior index $1 \le i \le n-2$ and every $t \in [t_i, t_{i+1}]$, we have $\pair_{a^{(i)}, b}(t) > 0$.
\end{lemma}

\begin{proof}
Let $a = a^{(i)}$ and write $\pair_{a,b}(t) = \sum_{k=0}^n c_k t^k$. By definition of $a^{(i)}$, we have $a_{i-1} = u_{i-1}$ and $a_{i+2} = u_{i+2}$, so Lemma~\ref{lem:tails}(i) gives $c_{i-1} \ge (1 - 4\epsilon) b_{i-1} u_{i-1} > 0$ and $c_{i+2} \ge (1 - 4\epsilon) b_{i+2} u_{i+2} > 0$. In the scaled function $g_i(t) \coloneqq t^{-i}\pair_{a,b}(t)$, these two terms carry powers $t^{-1}$ and $t^2$, whose second derivatives are strictly positive, while the central terms $c_i$ and $c_{i+1} t$ vanish upon differentiating twice. By Lemma~\ref{lem:tails}(i)--(ii), relative to the nearer of the two positive second derivatives ($2 c_{i-1} t^{-3}$ on the low side and $2 c_{i+2}$ on the high side), the sum of the absolute values of all more distant second derivatives on the same side is bounded by
\begin{equation}\label{eq:tailbudget}
  \frac{4}{1 - 4\epsilon} \sum_{h \ge 1} \frac{(h+1)(h+2)}{2}\,q_0^h
  = \frac{4}{1 - 4\epsilon}\bigl[(1 - q_0)^{-3} - 1\bigr] < \frac{1}{2}.
\end{equation}
Hence $g_i''(t) > 0$ throughout $(t_i, t_{i+1})$, so $g_i$ is strictly convex on $[t_i, t_{i+1}]$.
\begin{itemize}[leftmargin=1.8em,itemsep=2pt]
  \item If $1 < D_i \le 4$, then $g_i(T_i v) = b_i l_i H_i(v)$ attains a strictly positive stationary value at $T_i v_i \in (t_i, t_{i+1})$ by Proposition~\ref{prop:stationary} and Lemma~\ref{lem:bands}. Strict convexity implies $g_i(t) > 0$ on all of $[t_i, t_{i+1}]$.
  \item If $D_i > 4$, then $x_i = 1$ and $A_i = B_i = 1/\sqrt{D_i} < 1/2$. In $c_i$, the neighbor $b_{i+1} u_{i-1}$ costs $B_i < 1/2$ relative to the central term $b_i l_i$, while the other neighbor $b_{i-1} l_{i+1}$ uses the lower endpoint $l_{i+1}$ and costs at most $(8/7)\epsilon$. Symmetrically, in $c_{i+1}$, the neighbor $b_i u_{i+2}$ costs $A_i < 1/2$ relative to $b_{i+1} l_{i+1}$, and the other neighbor costs at most $(8/7)\epsilon$. Combining with the distance-$\ge 2$ bound $8\epsilon^2$ from Lemma~\ref{lem:tails}, both central coefficients satisfy
  \[
    \frac{c_i}{b_i l_i},\ \frac{c_{i+1}}{b_{i+1} l_{i+1}} \ge 1 - \frac{1}{2} - \frac{8}{7}\epsilon - 8\epsilon^2 > 0.
  \]
  Meanwhile, the estimate~\eqref{eq:tailbudget} (without the derivative factors) shows that the remaining tails ($k \le i-2$ and $k \ge i+3$) are bounded in magnitude by less than half of the two positive outer terms $c_{i-1}t^{i-1}$ and $c_{i+2}t^{i+2}$. Therefore $\pair_{a,b}(t) > 0$ on $[t_i, t_{i+1}]$. \qedhere
\end{itemize}
\end{proof}

\begin{lemma}[Positivity on the first and last bands]\label{lem:boundary-bands}
We have $\pair_{a^{(0)}, b}(t) > 0$ for all $t \in [0, t_1]$ and $\pair_{a^{(n-1)}, b}(t) > 0$ for all $t \in [t_{n-1}, \infty)$.
\end{lemma}

\begin{proof}
On band $0$, $a = a^{(0)}$ has $a_0 = l_0$, $a_1 = l_1$, and $a_2 = u_2$. The constant and linear coefficients of $\pair_{a,b}(t)$ are $c_0 = b_0 l_0 > 0$ and $c_1 = S - b_0 u_2 - M$, where $S \coloneqq b_1 l_1$ and $M \coloneqq b_2 l_0$. Subtracting the high tail ($k \ge 3$) from $c_2 t^2$ via Lemma~\ref{lem:tails} leaves at least $\frac{1}{2} b_2 u_2 t^2$ (since $1 - 4\epsilon - 4q_0/(1 - q_0) > 1/2$). By construction of $\beta_0$, $b_0 u_2 = \delta S$ with $\delta = 1/16$, while~\eqref{eq:betaineq} and Proposition~\ref{prop:stationary} give $M/S \le 1 + 8000\epsilon$. If $M \le (1 - 2\delta)S$, then $c_1 > 0$ and the quadratic $c_0 + c_1 t + \frac{1}{2} b_2 u_2 t^2$ is strictly positive on $[0, \infty)$. Otherwise $M > (1 - 2\delta)S$, and the negative part of $c_1$ satisfies $(-c_1)_+ \le (\delta + 8000\epsilon)S$; dividing its square by $c_0 \cdot (\frac{1}{2} b_2 u_2) = \frac{1}{2} b_0 u_2 M > \frac{1}{2}\delta(1 - 2\delta)S^2$ yields
\[
  \frac{(-c_1)_+^2}{\frac{1}{2} c_0 b_2 u_2}
  < \frac{2(\delta + 8000\epsilon)^2}{\delta(1 - 2\delta)} < 4.
\]
Hence if $c_1 \ge 0$, the quadratic $c_0 + c_1 t + \frac{1}{2} b_2 u_2 t^2$ is strictly positive on $[0, \infty)$ directly; if $c_1 < 0$, then $(-c_1)_+^2 = c_1^2$, so the bound above shows that its discriminant is strictly negative. In either case, $\pair_{a^{(0)}, b}(t) > 0$ for all $t \in [0, t_1]$.

On the last band $i = n-1$, reversing the coefficient order and substituting $t \mapsto 1/t$ yields the exact same quadratic estimate and sign-case split, since $b_n u_{n-2} = \delta b_{n-1} l_{n-1}$ and the reversed multiplier ratio is bounded by $1/(1 - 4000\epsilon) \le 1 + 8000\epsilon$.
\end{proof}

\begin{proof}[Proof of Theorem~\ref{thm:positive}]
For $n = 1$, $b(s) = 1 + s$ gives $\pair_{a,b}(t) = a_0 + a_1 t > 0$ for all $t \ge 0$. For $n = 2$, pick $b_1 = 1$ and choose $b_0, b_2 > 0$ small enough that $b_0 u_2 + b_2 u_0 < l_1$; then $\pair_{a,b}(t) = b_2 a_2 t^2 + (a_1 - b_0 a_2 - b_2 a_0)t + b_0 a_0$ has all three coefficients strictly positive across $\bx$.

For $n \ge 3$, take the degree-$n$ polynomial $b$ constructed in Proposition~\ref{prop:stationary} (whose leading coefficient $b_n > 0$). By Lemma~\ref{lem:bands}, the bands $[t_i, t_{i+1}]$ ($0 \le i < n$) cover $[0, \infty)$ with minimizing vertex $a^{(i)}$. Lemmas~\ref{lem:interior-bands} and~\ref{lem:boundary-bands} show that $\pair_{a^{(i)}, b}(t) > 0$ on each band, so $\pair_{a,b}(t) > 0$ for all $a \in \bx$ and $t \ge 0$ by Lemma~\ref{lem:four-vertex}. Equation~\eqref{eq:re-ratio} and conjugate symmetry then yield~\eqref{eq:target} for every $a \in \bx$ and every $\omega \in \R$.
\end{proof}

\section{Conclusion}\label{sec:conclusion}

We have shown that robust Hurwitz stability of an $n$th-order interval polynomial family does not guarantee equal-degree SPR synthesis in general: for every $n \ge 10$, there exist full-width rational Hurwitz boxes that exclude every nonzero real numerator of degree at most $n$, even under the weaker nonnegative real-part condition. Conversely, Theorem~\ref{thm:positive} proves that every obstruction must have at least one interior coefficient interval with $l_j/u_j > 10^{-5}$, since sufficiently wide interior variation decouples frequency scales and guarantees a common degree-$n$ SPR numerator in every order $n \ge 1$. Two natural open questions remain: whether equal-degree interval SPR synthesis holds for the intermediate orders $n \in \{7, 8, 9\}$, and what the sharp threshold for the interior endpoint ratio $l_j/u_j$ is.

\section*{Acknowledgements}
GPT-6 Astra was used for calculations, proof development, and editorial assistance. AI-generated suggestions were verified against the mathematical arguments and the accompanying formal proof; responsibility for the final manuscript rests with the authors. Both main theorems of this paper have been formally verified in Lean~4 (declarations \path{SPR.N10.Direct.counterexample} and \path{SPR.General.Direct.counterexample} for Theorem~\ref{thm:negative}, and \path{SPR.General.Direct.uniformWideSynthesis} for Theorem~\ref{thm:positive}, in \path{Lean/Main.lean}), and the source code repository is available at \url{https://github.com/dgjszy/spr-counterexamples}.

\appendix
\section{Rational Data for the Order-Ten Certificate}\label{app:certificate}

This appendix specifies the exact rational endpoints, root-bracketing tests, and twelve evaluation rows used in Proposition~\ref{prop:seed}. All operations are carried out over $\Q$ in the squared frequency variable $t = \omega^2$.

\subsection{Interval Endpoints and Root-Bracketing Signs}

Fix the rational parameters
\[
  z = 10^{-24}, \qquad e = z^{10}, \qquad \gamma = 10^{-6}, \qquad
  \nu = 10^{-8}, \qquad q = 1 - 10^{-300}.
\]
For $y \in \{q, 1\}$, define the eleven endpoint ratios
\[
  (\xi_0, \ldots, \xi_{10}) = (e,\ \gamma/y,\ y,\ y,\ e,\ \gamma/y,\ y,\ y,\ e,\ \gamma/y,\ 1).
\]
Set $d_h = f_h = (1 + \gamma)^{-1}$ for $h \in \{1, 3, 4\}$, $(d_2, f_2) = (1 - \gamma\nu, \nu)$, and $(g_{2h-1}, g_{2h}) = (1 - z^{h+1})(d_h, f_h)$ for $h = 1, 2, 3, 4$. Starting from $U_0 = U_1 = U_2 = 1$, compute recursively
\begin{equation}\label{eq:app-rec}
  U_{i+2} = g_i \xi_i \xi_{i+1} \frac{U_i U_{i+1}}{U_{i-1}} \quad (1 \le i \le 8), \qquad
  u_j(y) = \frac{U_j}{U_{10}}, \qquad l_j(y) = \xi_j u_j(y) \quad (0 \le j \le 10).
\end{equation}
Below, unadorned symbols $(l_j, u_j)$ denote the actual box endpoints ($y = q$, where $l_j < u_j$ for all $j \le 9$ and $l_{10} = u_{10} = 1$), whereas barred symbols $(\bar l_j, \bar u_j, \bar\xi_j)$ denote the auxiliary values at $y = 1$.

For $h = 1, 2, 3, 4$, define the reference frequencies and bracket endpoints
\[
  r_h = \frac{\bar u_{2h-1}}{\bar u_{2h+1}}, \qquad
  L_h = r_h \bar\xi_{2h-1}(1 - 10^{-155}), \qquad
  H_h = \frac{r_h(1 + 10^{-155})}{\bar\xi_{2h+1}},
\]
which satisfy $0 < L_1 < H_1 < L_2 < H_2 < L_3 < H_3 < L_4 < H_4$. Put $s_h = (-1)^{h-1}$. For each $h$, let the superscripts $+$ and $-$ denote the vertex choices maximizing and minimizing $s_h E_a(t)$ and $s_h O_a(t)$, respectively: specifically, maximizing takes $u_j$ when $s_h(-1)^{\lfloor j/2 \rfloor} > 0$ and $l_j$ otherwise on the corresponding parity, and minimizing reverses the endpoints. For example, $a_h^{+-}$ maximizes $s_h E_a$ and minimizes $s_h O_a$. Exact rational evaluation (via Horner's scheme: initialize $E = a_{10}$, $O = a_9$, and for $k = 4, 3, 2, 1, 0$ update $E \leftarrow a_{2k} - t E$ and, when $k < 4$, $O \leftarrow a_{2k+1} - t O$) verifies the sixteen strict inequalities
\[
  \begin{aligned}
    s_h E_{a_h^{+-}}(L_h) &< 0, & s_h O_{a_h^{+-}}(L_h) &> 0, \\
    s_h E_{a_h^{++}}(H_h) &< 0, & s_h O_{a_h^{++}}(H_h) &< 0
  \end{aligned}
  \qquad (h = 1, 2, 3, 4).
\]
By extremality of the four vertices (Lemma~\ref{lem:four-vertex}), these inequalities hold for every $a \in \bx_{10}$ and yield the root-bracketing and Hermite--Biehler interlacing used in Proposition~\ref{prop:seed}.

\subsection{The Twelve Test Rows and Rational Verification}

Set $\alpha = 10^{-15}$. The first eleven tests $(a_m, t_m)$ take $t_m = r_h v$ and the actual vertex of $\bx_{10}$ listed in Table~\ref{tab:tests} (using the oriented $+ / -$ convention for $s_h E_a, s_h O_a$ defined above).

\begin{table}[htbp]
\centering
\caption{The first eleven squared-frequency tests $(a_m, t_m)$ with $t_m = r_h v$.}\label{tab:tests}
\vspace{3pt}
\renewcommand{\arraystretch}{1.15}
\begin{tabular}{ccrc@{\hspace{14pt}}|@{\hspace{14pt}}ccrc}
\toprule
$m$ & $h$ & $v$ & vertex & $m$ & $h$ & $v$ & vertex \\
\midrule
$1$ & $1$ & $\gamma(1 - \alpha)$ & $+-$ & $7$ & $3$ & $\gamma(1 - \alpha)$ & $+-$ \\
$2$ & $1$ & $1 + \alpha$ & $++$ & $8$ & $3$ & $1 - \alpha$ & $++$ \\
$3$ & $1$ & $10^{-18}$ & $--$ & $9$ & $3$ & $1/4$ & $+-$ \\
$4$ & $1$ & $4$ & $-+$ & $10$ & $4$ & $1 + \alpha$ & $+-$ \\
$5$ & $2$ & $1 + \alpha$ & $+-$ & $11$ & $4$ & $(1 - \alpha)/\gamma$ & $++$ \\
$6$ & $2$ & $4$ & $++$ & & & & \\
\bottomrule
\end{tabular}
\end{table}

To specify the twelfth test, set $p_1 = 1$, compute recursively for $h = 1, 2, 3, 4$:
\[
  p_{2h+1} = \frac{p_{2h-1}}{r_h}, \qquad
  p_{2h} = p_{2h+1}\,\frac{\bar u_{2h-1}}{\bar l_{2h}},
\]
and put $p_0 = p_1 \bar l_1 / \bar u_2$ and $p_{10} = p_9 \bar l_9 / (e \bar u_8)$. Reuse the vertex $a_{12} = a_* \coloneqq a_1^{--}$ of row $3$ at the squared frequency
\[
  t_{12} = 10^{12}\left(1 + \sum_{j=0}^{9} (a_*)_j\right)\left(1 + \sum_{j=0}^{9} \frac{p_j}{p_{10}}\right).
\]

Form the $12 \times 11$ rational matrix $R$ with rows $R(a_m, t_m)$ from~\eqref{eq:row-entries}. Let $R_{\widehat{m}}$ denote the $11 \times 11$ matrix obtained by deleting row $m$, and set $d_m = (-1)^m \det R_{\widehat{m}}$ ($1 \le m \le 12$). Then~\eqref{eq:moments} is verified without any choice of normalization by the rational inequalities
\begin{equation}\label{eq:detcheck}
  d_{12} \ne 0, \qquad
  \frac{d_m}{d_{12}} > 0 \quad (1 \le m \le 11), \qquad
  \sum_{m=1}^{12} \frac{d_m}{d_{12}}\,O_{a_m}(t_m) > 0.
\end{equation}
Equivalently, solving the single $11 \times 11$ linear system $R_{\widehat{12}}^{\mathsf{T}} w = -R_{12}^{\mathsf{T}}$ over $\Q$, setting $w_{12} = 1$, checking $w_m > 0$ ($1 \le m \le 11$), and verifying $\sum_{m=1}^{12} w_m O_{a_m}(t_m) > 0$ establishes both assertions in~\eqref{eq:moments}. Positive row and column scalings may be applied during elimination provided any row scaling is applied to the $O_{a_m}(t_m)$ column as well.

\bibliographystyle{plain}
\bibliography{references}

\end{document}